\documentclass{ptephy}

\usepackage{amsmath} 
\usepackage{amsthm} 
\usepackage{cite} 
\usepackage{graphicx} 
\usepackage{algorithmic} 
\usepackage{subfig} 
\usepackage{url} 
\usepackage{latexsym}
\usepackage{natbib}
\bibliography{sample}

\begin{document}
\title{Cosmological entropy production and viscous processes
in the $(1+3+6)$-dimensional space-times}  
\author{\name{K. Tomita}{\ast}}
\address{\affil{}{Yukawa Institute for Theoretical Physics, Kyoto University, Kyoto 606-8502, Japan}
\rm{\email{ketomita@ybb.ne.jp}}}

\begin{abstract}
Cosmological entropy production is studied in the
$(1+3+6)$-dimensional space-times consisting of the outer space (the
$3$-dimensional expanding section) and the inner space (the
$6$-dimensional section). The inner space expands initially and contracts later.
First it is shown how the production of the $3$-dimensional entropy $S_3$
within the horizon is strengthened by the dissipation due to viscous
processes between the two spaces, in which we consider the viscosity 
caused by the gravitational-wave transport. Next it is shown under what
conditions we can have the critical epoch when $S_3$ reaches the
value $10^{88}$ in the Guth level and at the same time the outer space
is decoupled from the inner space.  Moreover, the total entropy $S_9$ in 
the $9$-dimensional space at the primeval expanding stage is also shown
 corresponding to $S_3$.  
\end{abstract}

\maketitle

\section{Introduction}
Our observed Universe consists of $4$-dimensional space-time and its
3-dimensional spatial section is very isotropic, homogeneous and
flat. According to the super-string theories, on the other hand, the
space-time originally has $10\ (=1+9)$ dimensions and our Universe
 is considered as a
partial section of the total space-time after its evolution. In order
that the section may be our observed Universe, it must satisfy the 
famous cosmological condition that it has the vast entropy $\sim
10^{88}$ within the horizon-size region.\citep{guth}

Recently the evolution of the space-time was analyzed by Kim et
al.\citep{kim1,kim2} in a matrix model of super-string theory and it was
shown that owing to the dimensional symmetry-breaking the total 
$9$-dimensional space is
separated into the outer space (the $3$-dimensional expanding section)
and the inner space (the $6$-dimensional expanding section), and that
the expansion 
rate of the inner space is smaller than that of the outer space. As
suggested by them, this may show 
the beginning of the separation of our $(1+3)$-dimensional Universe
from the 
other section. In order that our Universe may form in this direction,
the evolution of the inner space must tend from the expanding
phase to the contracting phase, collapse and finally decouple from the
expanding outer space, while the outer space inflates and tends to the
Friedman phase. A similar scenario of such a dynamic evolution 
of anisotropic multi-dimensional space-times was studied in the form
of Kaluza-Klein models in 1980-1990 \citep{chod,sahd,ishi,over}.
The cosmological entropy problem was also discussed in many
papers\citep{alv,abb1,abb2,barr,kolb,maeda,TI} in the framework of
classical relativity. Kolb et al.\citep{kolb} paid attention to the freeze-out 
epoch and found it is impossible, if the dimension of the inner space is
less than $16$, that at that epoch we obtain the 3-dimensional
entropy within the horizon in the Guth level and the outer
space is decoupled from the inner space. Abbott et al.\citep{abb1,abb2}
showed that it is possible, if the dimension of the inner space is
$\sim 40$.
 
In this paper we consider the entropy production which is obtained at an epoch
different from the freeze-out epoch, so as to avoid Kolb et al.'s and Abbott 
et al.'s above results. For this purpose we show in Sect. 2 our previous 
treatment\citep{TI} for the dynamics of multi-dimensional space-times. 
 As for the initial condition, the multi-dimensional universe is assumed to
 start from the state of nearly isotropic expansion, in the same way
 as Kolb et al.'s and Abbott et al.'s treatments, but in a different way from 
that in our previous one.\citep{TI} in which we treated only highly anisotropic 
cases. We include imperfect fluid with viscosity caused by the transport of 
gravitational waves, as well as the perfect fluid.
In Sect. 3 we describe the difference between 
the freeze-out epoch ($t_*$), the decoupling (or stabilization)
epoch and the epoch ($t_\dagger$) when the $3$-dimensional entropy
$S_3$ within the 
horizon reaches the critical value $10^{88}$.  In Sect. 4 we study first
the behaviors of $S_3$ and the total entropy $S$, and compare them
in two cases with non-viscous and viscous fluids. The role of
viscosity is found to be so strong as to bring vast entropies 
at the final stage in the collapse of the compact inner space and
the inflation of the outer space. Next, using approximate power solutions
(at the final stage), the condition (A) for 
$S_3 \sim 10^{88}$ is derived, together with the condition (B) that the
outer space 
should be decoupled from the inner space, and the compatibility of the
two conditions A and B at epoch $t_\dagger$ (different from $t_*$) is
shown.  

In Sect. 5, moreover, we solve numerically the differential evolution
equations for scale factors from the initial singular epoch to the final
 epoch, taking account of viscosity due to 
the gravitational-wave transport and derive the model parameters
satisfying the above two conditions A and B, by comparing their
solutions (at the later stage) with the asymptotic power solutions. 
In Sect. 6 we derive the (9-dimensional) primeval total entropy $S_9$, 
and it is found that the 
primeval entropy in the viscous case is much smaller than that in the
non-viscous case. In Sect. 7, the epoch which may cause the
dimensional symmetry-breaking is discussed from the viewpoint of
energy conservation, and the possible epoch of symmetry-breaking is
estimated. In Sect. 8 concluding remarks are given. 
In Appendix A the treatment of  
imperfect fluids is shown, in Appendix B the Planck length in the
outer space is derived in connection with the sizes of these two
spaces, and in Appendix C the relations between $S_3$ and $S_9$ are
derived.  
 
\section{Multi-dimensional space-times with viscous fluid}
First we assume that the fluid consists of massless particles with the
energy density $\epsilon$ and the pressure $p$ expressed as
\begin{equation}
  \label{eq:a1}
\epsilon = {\cal N} a_n T^{4+n} \quad {\rm and} \quad p = \epsilon/(3+n),
\end{equation}
where $T$ is the temperature, \ $n (= 6)$ is the dimension of the inner
space, \ $a_n$ is the $(4+n)$-dimensional Stefan-Boltzmann constant defined
by 
\begin{equation}
  \label{eq:a2}
a_n \equiv (3+n) \ \Gamma((4+n)/2) \ \zeta (4+n)/\pi^{(4+n)/2},
\end{equation}
and ${\cal N}$ is the number of particle species. The units $c = \hbar =
k$ (the Boltzmann constant) $= 1$ are used.

Since the total entropy $S$ within a comoving volume $V$ is given from
the second law of thermodynamics by
\begin{equation}
  \label{eq:a3}
TdS = V[d\epsilon + (\epsilon + p) dV/V],
\end{equation}
$S$ is expressed as
\begin{equation}
  \label{eq:a4}
S/V = [(4+n)/(2+n)] {\cal N} a_n (\epsilon/{\cal N} a_n)^{(3+n)/(4+n)}.
\end{equation}

As viscous quantities, we have the shear viscosity $\eta$ and the bulk
viscosity $\zeta$, but $\zeta$ vanishes for the fluid of massless particles. 
In imperfect fluids, propagating gravitational waves are absorbed in
multi-dimensional universes as well as in $4$-dimensional universes. The
corresponding $\eta$ is expressed as
\begin{equation}
  \label{eq:a5}
\kappa \eta = \eta_0 \sqrt{\kappa \epsilon},
\end{equation}
where
\begin{equation}
  \label{eq:a6}
\eta_0 \equiv \Bigl[\frac{4+n}{2(3+n)(5+n)}\frac{{\cal N}_r}{{\cal
N}}\Bigr]^{1/2}, 
\end{equation}
 $G = \kappa/8\pi$ is a $(4+n)$-dimensional gravitational constant,
and ${\cal 
N}_r$ is the number of radiative particle species absorbing
gravitational waves. We consider states so hot that there are many
interactions between particles, and so we assume ${\cal N}_r / {\cal
N} = 1$ for simplicity. 
The definition of $\eta$ and $\zeta$ and the derivation of
Eq.(\ref{eq:a6}) are shown in Appendix A. 

The space-times are described by the line element
\begin{equation}
  \label{eq:a7}
\begin{split}
ds^2 &= g_{MN} dx^M dx^N           \\
&=  -c^2 dt^2 + f(t)^2 g_{i j} dx^i dx^j +h(t)^2 g_{\alpha \beta}
dx^\alpha dx^\beta,
\end{split}
\end{equation}
where $M, N = 0, ..., (3+n), \quad i,j = 1, 2, 3,$ and $\alpha, \beta
= 4, ..., (3+n)$. The spaces with metrics $g_{ij}$ and $g_{\alpha
\beta}$ are the spaces with constant curvatures $k_f$ and  $k_h \
 (=0$ or $\pm 1)$. The Einstein equations are 
\begin{equation}
  \label{eq:a8}
R_{MN} - \frac{1}{2} R g_{MN} = \kappa T_{MN},
\end{equation}
where $G = \kappa /8\pi$ is a $(4+n)$-dimensional gravitational
constant. The energy-momentum tensor $T_{MN}$ for fluids with energy
density $\epsilon$, pressure $p$ and viscosity $\eta$ is
defined in Appendix A, and their components are
expressed under the comoving condition ($u^L = \delta^L_0$) as
\begin{equation}
  \label{eq:a9}
\begin{split}
T^0_0 &= -\epsilon, \quad T^0_M = 0 \quad (M \ne 0),  \\
T^M_N &= p' \delta^M_N - \eta \kappa^M_N \quad  (M,N \ne 0),
\end{split}
\end{equation}
where $\kappa^i_j = 2(\dot f/f)\delta^i_j, \kappa^\alpha_\beta =
2(\dot h/h)\delta^\alpha_\beta,$ \ (an overdot denoting
$\partial/\partial t$) and   
\begin{equation}
  \label{eq:a10}
p' \equiv p - \bigl[\zeta - \frac{2}{3+n}\eta\bigr] {\dot V}/V,
\end{equation}
where ${\dot V}/V = 3 {\dot f}/f + n {\dot h}/h$.
Then the above Einstein equations are expressed as
\begin{equation}
  \label{eq:a11}
\begin{split}
\ddot f/f &= -2({\dot f}^2 + k_f)/f^2 - n {\dot f} {\dot h} /(fh) -
2\kappa \eta {\dot f}/f + \kappa A/ (2+n), \\
\ddot h/h &= -(n-1)({\dot h}^2 + k_h)/h^2 - 3 {\dot f}{\dot h} /(fh) -
2\kappa \eta {\dot h}/h + \kappa A/ (2+n),
\end{split}
\end{equation}
where 
\begin{equation}
  \label{eq:a12}
A \equiv \epsilon - p + \bigl[\zeta + \frac{2(2+n)}{3+n} \eta\bigr]
{\dot V}/V, 
\end{equation}
and 
\begin{equation}
  \label{eq:a13}
\kappa \epsilon = 3({\dot f}^2 +k_f)/f^2 + \frac{1}{2}n(n-1) ({\dot
h}^2 +k_h)/h^2 + 3n{\dot f}{\dot h}/(fh).
\end{equation}
From the above equations we can get 
\begin{equation}
  \label{eq:a14}
\begin{split}
\dot \epsilon = &- (\epsilon +p){\dot V}/V + \bigl[\zeta +
\frac{2(2+n)}{3+n} \eta\bigr] ({\dot V}/V)^2 -2\eta \{3({\dot f}/f)(2{\dot f}/f+
n{\dot h}/h) \\
 &+n({\dot h}/h)[3{\dot f}/f +(n-1){\dot h}/h]\}. 
\end{split}
\end{equation}
For the change in the total entropy $S$, we obtain from
Eqs.(\ref{eq:a4}) and (\ref{eq:a14})
\begin{equation}
  \label{eq:a15}
(\epsilon+p) {\dot S}/S = \zeta ({\dot V}/V)^2 + \frac{6n}{(3+n)} \eta 
({\dot f}/f - {\dot h}/h)^2 > 0.
\end{equation}
Evidently $S =$ const for the non-viscous case ($\eta = \zeta = 0$).

It is assumed that initially the universe expands isotropically, but the
expansion of the $n$-dimensional inner space is slower than that
of the 
$3$-dimensional outer space. At the later anisotropic stage the inner
space 
contracts, while the outer space continues to expand. These behaviors
correspond to the solutions with $k_h = 1$ and $k_f = 0$ or $-1$. Their
solutions can be derived solving Eq.(\ref{eq:a11}) numerically, but, paying 
attention to their
early isotropic stage and the later anisotropic stage (after the epoch
of maximum expansion of the inner space), we can use approximate power
solutions which are derived in the following, neglecting the curvature
terms. 
 They are expressed as   
\begin{equation}
  \label{eq:a16}
h = h_I (t - t_I)^\mu , \quad f = f_I (t - t_I)^\nu
\end{equation}
with $\mu = \nu$ at the isotropic stage, and 
\begin{equation}
  \label{eq:a17}
h = h_A (t_A - t)^\mu , \quad f = f_A (t_A - t)^\nu
\end{equation}
with $\mu \ne \nu$ at the anisotropic stage, where $h_I, f_I, t_I, h_A,
f_A,$ and $t_A$ are constants. The times $t_I$ and $t_A$ represent the
initial epoch of the isotropic stage and the final epoch of the
anisotropic stage, respectively. From Eq. (\ref{eq:a11}), we obtain the
solutions expressed as
\begin{equation}
  \label{eq:a18}
\begin{split}
(n \mu +3 \nu -1)(n \mu +3 \nu) - H &= 0,    \\
(\mu -\nu)(n \mu +3 \nu -1 -2\eta_0 \sqrt{H}) &= 0,
\end{split}
\end{equation}
where $H \equiv 3\nu^2 + \frac{1}{2}n(n-1)\mu^2 +3n\mu\nu$.

In the isotropic case, the solutions of these equations and the energy
density are
\begin{equation}
  \label{eq:a19}
\mu = \nu = 2/(4+n), \qquad \kappa \epsilon = H(t-t_I)^{-2} \ (> 0).
\end{equation}
In the anisotropic case, which we consider from now on, they are
\begin{equation}
  \label{eq:a20}
\begin{split}
\mu &= (1+\xi/n)[(3+n)(1-4{\eta_0}^2)]^{-1},   \\
\nu &= (1-\xi/3)[(3+n)(1-4{\eta_0}^2)]^{-1}
\end{split}
\end{equation}
with
\begin{equation}
  \label{eq:a21}
\xi^2 \equiv n^2 + 2n(3+n)(1 - 12{\eta_0}^2),
\end{equation}
and
\begin{equation}
  \label{eq:a22}
\kappa \epsilon = H (t_A - t)^{-2}, 
\end{equation}
where $H = 4 {\eta_0}^2/(1-4{\eta_0}^2)$. 
For viscous fluid and $n = 6$, we have $\mu = 0.345, \nu =
-0.272,$ and $\eta_0 = 0.225$, where Eq.(\ref{eq:a6}) was
used. Moreover, $H  > 0$ and so Eq.(\ref{eq:a22}) is applicable. 
For non-viscous fluid ($\eta_0 = 0$), we have $\mu = -\nu=
1/3$, but $H = 0$, so that Eq.(\ref{eq:a22}) is not applicable, and we
must take account of higher terms in $f$ and $h$ due to the 
curvature terms, to derive $\epsilon$. For $\eta_0 = 0$, on the other
hand, $S$ is constant and so from Eq.(\ref{eq:a4})
\begin{equation}
  \label{eq:a22a}
\epsilon \propto V^{-(4+n)/(3+n)} \propto (t_A -t)^{-\alpha},
\end{equation}
where $\alpha \equiv (n\mu +3\nu)(4+n)/(3+n) = 10/9$ for $n = 6$.

The temperature $T$ \ ($\propto \epsilon^{1/(n+4)}$) is expressed as  
\begin{equation}
  \label{eq:a23}
T \propto (t_A - t)^{-1/5}, \quad (t_A - t)^{-1/9} 
\end{equation}
for $\eta_0 \ne 0, = 0$, respectively. Accordingly, the temperature
change for $\eta_0 \ne 0$ is very rapid, compared with that for $\eta_0 =
0$. 

Around the epoch when the inner space takes maximum expansion, the above
power solution cannot be used and curvature terms must be
considered. Here we specify the epoch $t_M$ as the 
epoch which is comparatively near the epoch of maximum expansion but
when the power solutions are approximately applicable. In Sect.4 we consider
the anisotropic stage after $t_M$ for simplicity and use the approximate power
solutions.  In Sect. 5, 6 and 7, we treat the
solutions applicable at all epochs and compare them with the approximate 
power solutions.

\section{Epochs of freeze-out and decoupling of the inner space}
While the inner space continues to collapse, the outer space
inflates. In order that the expansion of the outer space may change to
the Friedman-like slow one, the inner space must be
separated from the outer space (see Fig. 1). This epoch is called the decoupling (or
stabilization) epoch. At this epoch the size of the inner space is
estimated to be comparable with or smaller than the Planck length in
the outer space, so that the space-time may be completely quantized on 
that scale in the outer space.

Around this epoch, we may have an epoch when the physical state
in the inner space may change.  Kolb et al. \citep{kolb} suggested that
classical treatments of space-time and fluids in the inner space may
be questionable around the epoch ($t_*$) when $h T = 1$, in which $h$
represents the size of the inner space and $T$ is related to
$\epsilon$ by $\epsilon = T^{4+n} \ (n = 6)$. This is because $h$ is
comparable with the mean wavelength of massless particles. This epoch
($t_*$) is called the freeze-out epoch.  Kolb et al. assumed
 that $t_*$ is near the decoupling epoch, and derived the
$3$-dimensional entropy $S_3$ within the horizon. But they found it is
impossible for it to reach the expected value $\sim 10^{88}$ in the
Guth level,\citep{guth} if the dimension $n$ of the inner space is less
than $16$. Abbott et al.\citep{abb1,abb2} found that at $t_*$ it
reaches the expected value, if $n$ is $\sim 40$. 

In the next section we analyze the entropies at $t_\dagger$ (which is
different from $t_*$), and derive the condition (A) for $S_3 = 10^{88}$  
 using classical relativity. Moreover, 
 we also consider  the Planck length in the outer
space ($f_{pl}$), corresponding to the Newtonian gravitational
constant ($G_N$). It is defined in the connection with $G \
(=\kappa/8\pi)$ by
\begin{equation}
  \label{eq:d1}
\kappa/8\pi = [h(t_\dagger)]^n {f_{pl}}^2.
\end{equation}
Using this relation, we derive the condition (B) that the size of the
inner space ${h(t)}$ is comparable with or smaller than $f_{pl}$, and
examine the compatibility between conditions A and B.

\begin{figure}[t]
\caption{\label{fig:entropy} Scale factors of outer and inner spaces.}
\includegraphics[width=15cm]{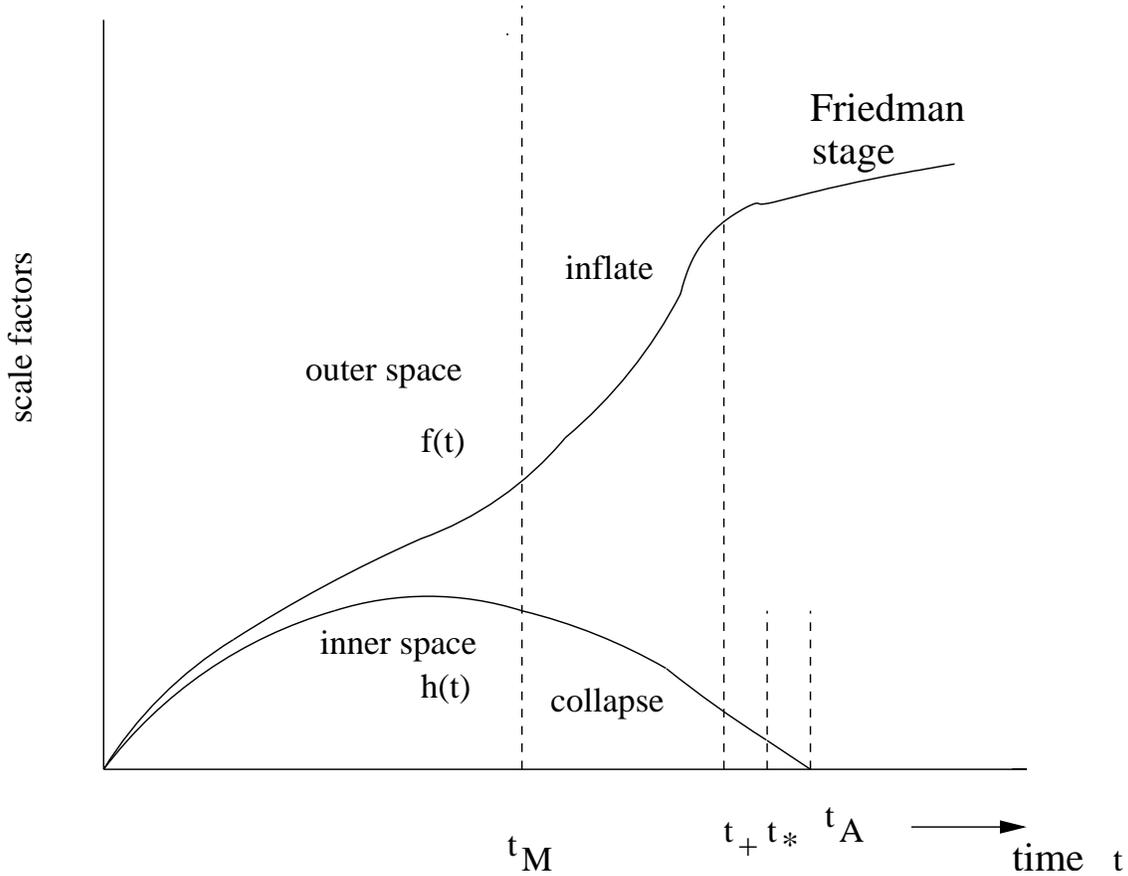}
\end{figure}

\section{Total entropy $S$ and the $3$-dimensional entropy $S_3$}
The total entropy $S$ in the multi-dimensional space-time is expressed
at the stage of $(t_A -t) \ll t_A$ using Eqs. (\ref{eq:a4}) and
(\ref{eq:a17}) as
\begin{equation}
  \label{eq:b1}
S \propto (t_A - t)^{-\beta},
\end{equation}
where $\beta \equiv (3+n)/(4+n) - (n\mu + 3\nu)$ for $\eta_0 \ne 0$.
For the non-viscous case, $S$ is constant.
The behavior of $S$ was discussed in Ref.\citep{TI} using the
inflation and collapse factors of the outer and inner spaces. 
The important quantity which is to be noticed directly from the
viewpoint of cosmological observations, however, is not $S$, but the
$3$-dimensional entropy $S_3$ within the horizon $l_h$ of
$3$-dimensional outer space. 

Here $S_3$ is defined by
\begin{equation}
  \label{eq:b2}
S_3 = s_3 (l_h)^3,
\end{equation}
where the $3$-dimensional entropy density $s_3$ is given by $s_3 =
{\epsilon_3}^{3/4}$, and the $3$-dimensional energy density $\epsilon_3$ is
\begin{equation}
  \label{eq:b3}
\epsilon_3 = h(t)^n \epsilon = h(t)^n T^{n+4}.
\end{equation}
In contrast to $S$, $S_3$ increases with time, not only in the viscous
case but also in the non-viscous case, because the common temperature $T$
 increases with the decrease of the volume $V$ of the total space in
both cases.   

Here and in the following we neglect the factors $\approx 1$ such as
$a_0$. Then  
\begin{equation}
  \label{eq:b4}
(S_3)^{1/3} = [h(t) T]^{\frac{n+4}{4}} h(t)^{-1} l_h,
\end{equation}
where the horizon-size $l_h$ is
\begin{equation}
  \label{eq:b5}
l_h = f(t) \int^t_0  dt'/f(t').
\end{equation}

\subsection{Entropy in the freeze-out epoch $t_*$}
At epoch $t_*$ with $h(t_*) T_* \equiv 1$, we have
\begin{equation}
  \label{eq:b6}
(S_3)_* = [h(t_*)^{-1} f(t_*)]^3 \Bigl[\int^{t_*}_0 dt'/f(t')\Bigr]^3
\sim [(t_A - t_*)/(t_A - t_M)]^{3(\nu-\mu)} {(f/h)_M}^3,
\end{equation}
where we assumed that this epoch $t_*$ is at the stage of 
$(t_A - t) \ll t_A$.

In the non-viscous case, we have $\mu= -\nu = 1/3$, and 
\begin{equation}
  \label{eq:b7}
(S_3)_* \sim [(t_A -t_*)/(t_A - t_M)]^{-2} {(f/h)_M}^3 =
[h(t_*)/h(t_M)]^{-6} {(f/h)_M}^3,
\end{equation}
which is consistent with Kolb et al.'s result (Eq. (5.5) in Ref.\citep{kolb}). 

In our viscous case ($n = 6$), we have $\mu = 0.345,  \nu = -0.272$ and
$\eta_0 = 0.225$ \ (cf. Eq.(\ref{eq:a6})), and 
\begin{equation}
  \label{eq:b8}
(S_3)_* \sim [(t_A -t_{*v})/(t_A - t_M)]^{-1.83} {(f/h)_M}^3 =
[h(t_{*v})/h(t_M)]^{-5.38} {(f/h)_M}^3 , 
\end{equation}
where we used the notation $t_{*v}$ to discriminate $t_*$ in the viscous
case from $t_*$ in the non-viscous case. 

Now let us compare quantities at $t_*$ and $t_{*v}$. Using
Eq.(\ref{eq:a23}), we have
\begin{equation}
  \label{eq:b9}
\begin{split}
\frac{h(t_*) T(t_*)}{h(t_M) T(t_M)} &= \Bigl(\frac{t_A - t_*}{t_A -
t_M}\Bigr)^{2/9},   \\ 
\frac{h(t_{*v}) T(t_{*v})}{h(t_M) T(t_M)} &= \Bigl(\frac{t_A -
t_{*v}}{t_A - t_M}\Bigr)^{0.145} 
\end{split}
\end{equation}
for $\eta_0 = 0, \ \eta_0 \ne 0$, respectively. Since $h(t_*) T(t_*)
= h(t_{*v}) T(t_{*v}) = 1$ and  $h(t_M) T(t_M)$ is common, we obtain
from these equations
\begin{equation}
  \label{eq:b10}
\Bigl(\frac{t_A - t_*}{t_A -t_M}\Bigr)^{2/9} =  \Bigl(\frac{t_A -
t_{*v}}{t_A - t_M}\Bigr)^{0.145},
\end{equation}
so that $S_3$ in the non-viscous case is expressed in terms of
$t_{*v}$ as
\begin{equation}
  \label{eq:b11}
(S_3)_* \sim \Bigl(\frac{t_A - t_{*v}}{t_A -t_M}\Bigr)^{0.131}
{(f/h)_M}^3 = [h(t_{*v})/h(t_M)]^{-3.78} {(f/h)_M}^3.
\end{equation}
By comparing this with Eq.(\ref{eq:b8}), therefore, it is found that
$(S_3)_*$ in the viscous case is much larger than $(S_3)_*$ in the
non-viscous case, since $h(t_{*v})/h(t_M) \ll 1$.

\subsection{Entropy at epoch $t_\dagger$ (different from $t_*$)}
Now let us derive the conditions A and B suggested in Sect. 3.
First we derive $S_3$ at epoch $t_\dagger$ with 
\begin{equation}
  \label{eq:b12}
h(t_\dagger) T_\dagger \equiv \lambda \ (\ne 1).
\end{equation}
In this case we have
\begin{equation}
  \label{eq:b13}
(S_3)^{1/3} = \lambda^{\frac{n+4}{4}} [h(t_\dagger)^{-1} f(t_\dagger)]
\int^{t_\dagger}_0 dt'/f(t').
\end{equation}
\noindent (a) the viscous case ($\eta_0 = 0.225$)

Using Eq.(\ref{eq:a23}) we obtain
\begin{equation}
  \label{eq:b14}
\lambda = [h(t_\dagger)/h(t_M)]^{1-\frac{1}{5 \mu}} h(t_M) T_M
\end{equation}
with $\mu = 0.345$.

For $S_3$ at $t_\dagger$, we obtain neglecting
the factors of integrals
\begin{equation}
  \label{eq:b16}
(S_3)_\dagger \sim [h(t_\dagger)/h(t_M)]^\gamma S_0
\end{equation}
with $\gamma \equiv -3(1 -\nu/\mu) =-5.362$ , where 
\begin{equation}
  \label{eq:b17}
S_0 \equiv \lambda^{3(n+4)/4} {(f/h)_M}^3
\end{equation}
and we assumed that the epoch $t_\dagger$ is at the stage of 
$(t_A - t) \ll t_A$. If we express $(S_3)_\dagger$ in terms of
 $\lambda$, we have by use of Eq.(\ref{eq:b14})  
\begin{equation}
  \label{eq:b18}
(S_3)_\dagger \sim \Bigl[{\lambda/(h T)_M}\Bigr]^\delta S_0
\end{equation}
with $\delta \equiv [\gamma/(1-\frac{1}{5\mu})] = -12.77$. Assuming
$(S_3)_\dagger = 10^{88}$, we have 
\begin{equation}
  \label{eq:b19}
\begin{split}
h(t_\dagger)/h(t_M) &= [(S_3)_\dagger/S_0]^{1/\gamma} = 10^{-16.4}
(S_0)^{0.186},   \\
{\lambda/ (h T)_M} &= [(S_3)_\dagger/S_0]^{1/\delta} = 10^{-6.89}
(S_0)^{0.0783}.  
\end{split}
\end{equation}
Moreover, we obtain from these relations
\begin{equation}
  \label{eq:b20}
T_\dagger /T_M = [\lambda/(hT)_M] [h(t_\dagger)/h(t_M)]^{-1} =
10^{9.51} (S_0)^{-0.108}. 
\end{equation}

Next we consider the relations of $f$ and $h$ to the Planck
length $f_{pl}$ in the outer space (cf. Sect. 3), which are derived in Appendix B.
In order that the space-time in the outer space can be treated
classically at epoch $t_\dagger$, the condition $f_\dagger/f_{pl}
\gg 1$ must be satisfied. This condition gives
$\lambda > 1.96$. Moreover, $h_\dagger/f_{pl}$ is smaller than $1$,
if $\lambda < 1.06\times 10^4$. It is found, therefore, that, if
$\lambda \sim 1.06\times 10^4$, the outer space at epoch
$t_\dagger$ gets $(S_3)_\dagger 
= 10^{88}$ and is decoupled from the inner space at the same time,
because it has the quantized space-time.

For $\lambda = 1.06\times 10^4$, we obtain  assuming $(f/h)_M = 1.5$ 
\begin{equation}
  \label{eq:b21}
S_0 = 5.22 \times 10^{30}
\end{equation}
from Eq.(\ref{eq:b17}),
\begin{equation}
  \label{eq:b22}
\begin{split}
h(t_\dagger)/h(t_M) &= 2.06\times 10^{-11},
  \\
{\lambda/ (h T)_M} &= 2.28\times 10^{-5}
\end{split}
\end{equation}
from Eq.(\ref{eq:b19}), and 
\begin{equation}
  \label{eq:b23}
T_\dagger /T_M = 1.59\times 10^{6}
\end{equation}
from Eq.(\ref{eq:b20}). The model-dependent value of $(f/h)_M$ 
is shown in Sect. 6. Since $f_{pl}$ corresponds to the Planck
temperature $T_{pl}$ \ ($= 10^{19}$ Gev) and $h_\dagger/f_{pl} = 1$,  
we have $T_\dagger = \lambda
T_{pl}$ and $T_M = 6.67\times 10^{-3} T_{pl}$ .
From Eq.(\ref{eq:b22}), moreover, we can also obtain 
\begin{equation}
  \label{eq:b24}
(h T)_M = 3.23\times 10^{8}.
\end{equation}

\noindent (b) the non-viscous case ($\eta_0 = 0$)

Using Eq.(\ref{eq:a23}) we obtain
\begin{equation}
  \label{eq:b15b}
\lambda = [h(t_\dagger)/h(t_M)]^{2/3} h(t_M) T_M.
\end{equation}
For $S_3$ at $t_\dagger$, we obtain neglecting the factors of integrals
\begin{equation}
  \label{eq:b16b}
(S_3)_\dagger \sim [h(t_\dagger)/h(t_M)]^\gamma S_0
\end{equation}
with $\gamma \equiv -3(1 -\nu/\mu) = -6$ and $S_0$ in Eq.(\ref{eq:b17}).
If we compare them in terms of $\lambda$, we have by use of
Eq. (\ref{eq:b15b}) 
\begin{equation}
  \label{eq:b18b}
(S_3)_\dagger \sim \Bigl[{\lambda/ (h T)_M}\Bigr]^\delta S_0
\end{equation}
with $\delta \equiv (3/2)\gamma = -9$. Assuming
$(S_3)_\dagger = 10^{88}$, we have 
\begin{equation}
  \label{eq:b19b}
\begin{split}
h(t_\dagger)/h(t_M) &= [(S_3)_\dagger/S_0]^{1/\gamma} = 10^{-14.7}
(S_0)^{0.167},  \\ 
{\lambda/ (h T)_M} &= [(S_3)_\dagger/S_0]^{1/\delta} = 10^{-9.78}
(S_0)^{0.111}.    
\end{split}
\end{equation}
Moreover, we obtain from these relations
\begin{equation}
  \label{eq:b20b}
T_\dagger /T_M = [\lambda/(hT)_M] [h(t_\dagger)/h(t_M)]^{-1} =
 10^{4.89} (S_0)^{-0.0556}. 
\end{equation}

Next we consider the relations of $f$ and $h$ to the Planck
length $f_{pl}$ in the outer space, which are derived in Appendix B.
In order that the space-time in the outer space can be treated
classically at epoch $t_\dagger$, the condition $f_\dagger/f_{pl}
\gg 1$ must be satisfied. This condition gives gives $\lambda >
10^{-5.88}$. Moreover, $h_\dagger/f_{pl}$ is smaller than $1$,
if $\lambda <  46.1$. It is found, therefore, that, if
$\lambda \sim 46.1$, the outer space at epoch
$t_\dagger$ gets $(S_3)_\dagger 
= 10^{88}$ and is decoupled from the inner space at the same time,
because it has the quantized space-time.

For $\lambda =  46.1$, we obtain  assuming $(f/h)_M = 1.5$ 
\begin{equation}
  \label{eq:b21b}
S_0 =  1.01 \times 10^{13}
\end{equation}
from Eq.(\ref{eq:b17}),
\begin{equation}
  \label{eq:b22b}
\begin{split}
h(t_\dagger)/h(t_M) &=  2.94\times 10^{-13},
  \\
{\lambda/ (h T)_M} &=  4.57\times 10^{-9}
\end{split}
\end{equation}
from Eq.(\ref{eq:b19b}), and 
\begin{equation}
  \label{eq:b23b}
T_\dagger /T_M = 1.55\times 10^{4}
\end{equation}
from Eq.(\ref{eq:b20b}).  Since $f_{pl}$ corresponds to the Planck
temperature $T_{pl}$ \ ($= 10^{19}$ Gev) and $h_\dagger/f_{pl} = 1$, 
we have $T_\dagger = \lambda
T_{pl}$ and $T_M = 2.97\times 10^{-3} T_{pl}$ .
From Eq.(\ref{eq:b22b}), moreover, we can obtain also
\begin{equation}
  \label{eq:b24b}
(h T)_M = 1.01\times 10^{10}.
\end{equation}

These values of $T_M$ and $(h T)_M $ give the condition on the thermal
state that the 
multi-dimensional universe must satisfy at the epoch near the maximum
expansion and so at the stage of the early isotropic expansion. 

\section{Numerical histories of multi-dimensional universes}
In this section let us solve numerically equations of scale factors
$f(t)$ and $h(t)$ in the outer and inner 
spaces and equations of the energy density $\epsilon (t)$ (given
by Eqs. (\ref{eq:a11}) and (\ref{eq:a13})) at the total stage and
relate their behaviors at the final 
asymptotic stage (which were treated in the previous section) to the
behaviors at the earliest stage.  
Here we assume that the inner space has positive curvature ($k_h =
1$) and the outer space is flat or has negative curvature ($k_f = 0$
or $-1$). 

Now let us transform variables $(t, f, h, \epsilon, p, \eta)$ to
$(\bar{t}, \bar{f}, \bar{h}, \bar{\epsilon}, \bar{p}, \bar{\eta})$ as follows, 
for the convenience of numerical calculations: 
\begin{equation}
  \label{eq:aa0}
t=\zeta \bar{t},\ f=\zeta \bar{f},\ h=\zeta \bar{h},\ \epsilon =\zeta^{-2} \bar{\epsilon}, \
p = \zeta^{-2} \bar{p},\ \eta= \zeta^{-1} \bar{\eta},
\end{equation}
where  $\zeta$ is a positive constant and $\eta_0, k_f, k_h$ are assumed to be invariant.
For this transformation, the forms of Eqs. (\ref{eq:a11}),  (\ref{eq:a12}),  and
(\ref{eq:a13}) are invariant. In this section these equations for the new 
variables are solved, where $\zeta$ is determined so as to be consistent with
the initial condition in the following.

To determine the initial condition at $t=t_i$ for solving these
equations, we consider the approximate solutions $f(t)$ and $h(t)$
around the isotropic solution $t^{1/5}$ (in the limit of small $t$) as 
\begin{equation}
  \label{eq:aa1}
\begin{split}
f &= f_0 t^{1/5} [1 + f_1(t)], \\
h &= h_0 t^{1/5} [1 + h_1(t)]
\end{split}
\end{equation}
with constants $h_0$ and $f_0 (= h_0)$, where we assume $f_1 \ll 1, \ \
h_1 \ll 1$. Here the isotropic solution is equal to Eqs. (\ref{eq:a16}) and
(\ref{eq:a19}) with $T_I = 0$, and $t_i$ is a small positive time.
Then we obtain from Eq.(\ref{eq:a11}) 
\begin{equation}
  \label{eq:aa2}
\begin{split}
\ddot{f_1} &= -\frac{28 +24\eta_0}{15t}\dot{f}_1 - \frac{2
-24\eta_0}{15t}\dot{h}_1 + \frac{5}{3} (1 -k_f)\frac{t^{-2/5}}{{h_0}^2},
\\  
\ddot{h_1} &= -\frac{1 -12\eta_0}{15t}\dot{f}_1 - \frac{29
+12\eta_0}{15t}\dot{h}_1 - \frac{(10 - k_f)}{3}
\frac{t^{-2/5}}{{h_0}^2},  
\end{split}
\end{equation}
where the last terms come from the positive curvature ($k_h = 1$) in
the inner space and the curvature $k_f (= 0$ or $1)$ in the outer
space, and from Eq.(\ref{eq:a13}) 
\begin{equation}
  \label{eq:aa3}
\kappa \epsilon = \frac{36}{25t^2}[1+\frac{10}{3}t(\dot{f}_1+
2\dot{h}_1) + 
\frac{25}{12}(5+k_f) \frac{t^{8/5}}{{h_0}^2}].
\end{equation}
Solving these equations, we obtain the following approximate solutions
\begin{equation}
  \label{eq:aa4}
\begin{split}
f &= h_0 t^{1/5} (1 + f_{10}t^{8/5}), \\
h &= h_0 t^{1/5} (1 + h_{10}t^{8/5}),
\end{split}
\end{equation}
where
\begin{equation}
  \label{eq:aa5}
\begin{split}
f_{10} &= \frac{125}{288\times 13}\frac{{14-12\eta_0
-\frac{4}{5}(16+3\eta_0) 
k_f}}{1+\eta_0} {h_0}^{-2}, \\
h_{10} &= -\frac{125}{288\times 13}\frac{{25+12\eta_0
-\frac{2}{5}(7-6\eta_0) k_f}}{1+\eta_0} {h_0}^{-2}, 
\end{split}
\end{equation}
and
\begin{equation}
  \label{eq:aa5a}
f_{10} -h_{10} =  \frac{125\times 3}{288} \frac{1- 2k_f}{1+\eta_0}
{h_0}^{-2}. 
\end{equation}

Next, using these solutions, we make $6$ types of initial conditions $f, h, 
\dot{f},$ and $\dot{h}$ at $t=t_i$ which are discriminated with
$\eta_0, h_0$ and $t_i$ as

a. $(010)\ : \quad \eta_0 = 0, \qquad \ \ h_0 = 1, \ \quad t_i = 0.1, $

b. $(007)\ : \quad \eta_0 = 0, \qquad \ \ h_0 = 0.7, \quad t_i = 0.064,$

c. $(015)\ : \quad \eta_0 = 0, \qquad \ \ h_0 = 1.5, \quad t_i = 0.166,$

d. $(210)\ : \quad \eta_0 = 0.225, \quad h_0 = 1, \ \quad t_i = 0.1, $

e. $(207)\ : \quad \eta_0 = 0.225, \quad h_0 = 0.7, \quad t_i = 0.064,$

f. $(215)\ : \quad \eta_0 = 0.225, \quad h_0 = 1.5, \quad t_i = 0.166,$

\noindent where these parameters satisfy the relation ${h_0}^{-2}
(t_i)^{8/5} = 0.025$.  After the transformation (\ref{eq:aa0}) corresponding to 
these initial conditions, $t$ is dimensionless, while $\zeta$ has a dimension of length.  

For these initial values, Eq.(\ref{eq:a11}) was solved numerically using the
Runge-Kutta method, and the result is shown by solid curves for $t
\geq t_i$ and by short-dashed curves for $t < t_i$ in
Figs. 2, 3, 4, 5, 6, and 7 for $k_f = 0$. The behavior of $f$ and $h$ for
$k_f = -1$ is similar to that for $k_f = 0$.

In these solutions we estimated the singular epoch $t= t_A$ and 
derived the power solutions with $f = f_A (t_A-t)^\nu, \ h = h_A
(t_A -t)^\mu$ using it, where $(\mu,\nu) = (1/3, -1/3), \ (0.345, 
-0.272)$ for
$\eta_0 = (0, \ 0.225)$, respectively.  Constants $f_A$ and $h_A$ are
determined so that these power solutions are tangent to the true
solutions at epoch $t_A$.  These solutions are shown by  
long-dashed curves in Figs. 2, 3, 4, 5, 6, and 7.   

Moreover, the factors [appearing in Eqs.(\ref{eq:e2}) and
(\ref{eq:e5}))]   
\begin{equation}
  \label{eq:aa6}
 \begin{split}
\Phi_0 (t) &\equiv  \sqrt{\frac{8\pi}{\kappa \epsilon}} (f/h^2) (h/f)^{0.667},\\
\Phi_2 (t) &\equiv  \sqrt{\frac{8\pi}{\kappa \epsilon}} (f/h^2) (h/f)^{0.06}
\end{split}
\end{equation}
for $\eta_0 = 0, \ 0.225$, respectively, are shown as functions of $t$
by solid curves in Figs. 8, 9, 10, 11, 12, and 13 for $k_f = 0$. The temperature 
$T \ [= (\epsilon/{\cal N} 
a_n)^{1/10}]$ is also shown by dashed curves in the same figures.
These two quantities $\Phi_0 (t)$ and $\Phi_2 (t)$ are invariant for the
transformation (\ref{eq:aa0}) and so they can be regarded as quantities 
on the ordinary scale of $(t, f, h, \epsilon, p, \eta)$. Moreover, $\lambda, S_0$
and the ratios of $f, h,$ and $T$  
also, which are determined in the next section using these $\Phi_0 (t)$ and 
$\Phi_2 (t)$, can be treated as quantities on the ordinary scale. 
\begin{figure}[t]
\caption{\label{fig:1a} Scale factors of the outer and inner
spaces in a.(010). Solid and short-dashed curves denote the scale
factors $f$ and $h$ for $t \geq t_i$ and $< t_i$, respectively. 
Long-dashed curves denote the power solutions tangent to $f$ and $h$ at
epoch $t_A$. The thin solid line denotes the possible epoch of 
dimensional symmetry-breaking.} 
\centerline{\includegraphics[width=8cm]{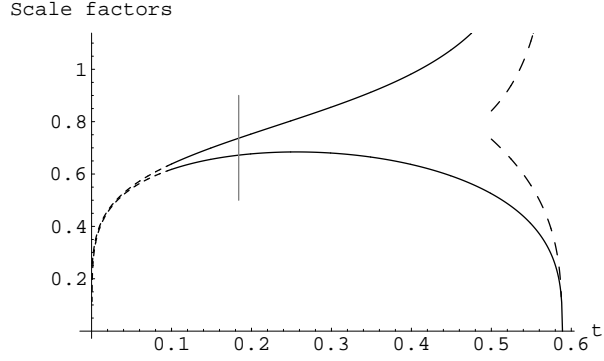}}
\end{figure}
\begin{figure}[t]
\caption{\label{fig:1b} Scale factors of the outer and inner
spaces in b.(007). The notation of the curves is  as in Fig. 2.} 
\centerline{\includegraphics[width=8cm]{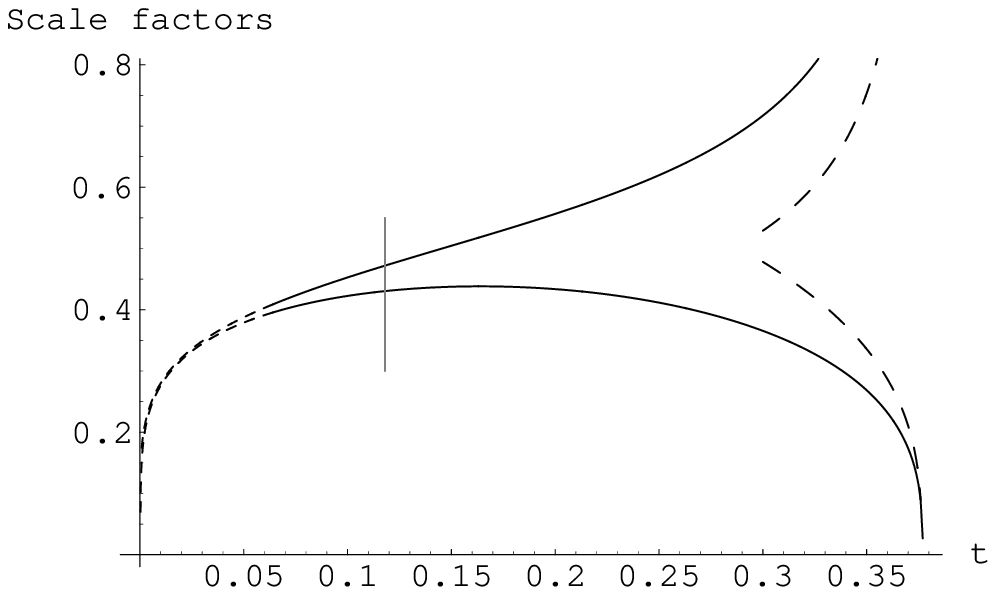}}
\end{figure}
\begin{figure}[t]
\caption{\label{fig:1c} Scale factors of the outer and inner
spaces in c.(015). The notation of the curves is  as in Fig. 2.} 
\centerline{\includegraphics[width=8cm]{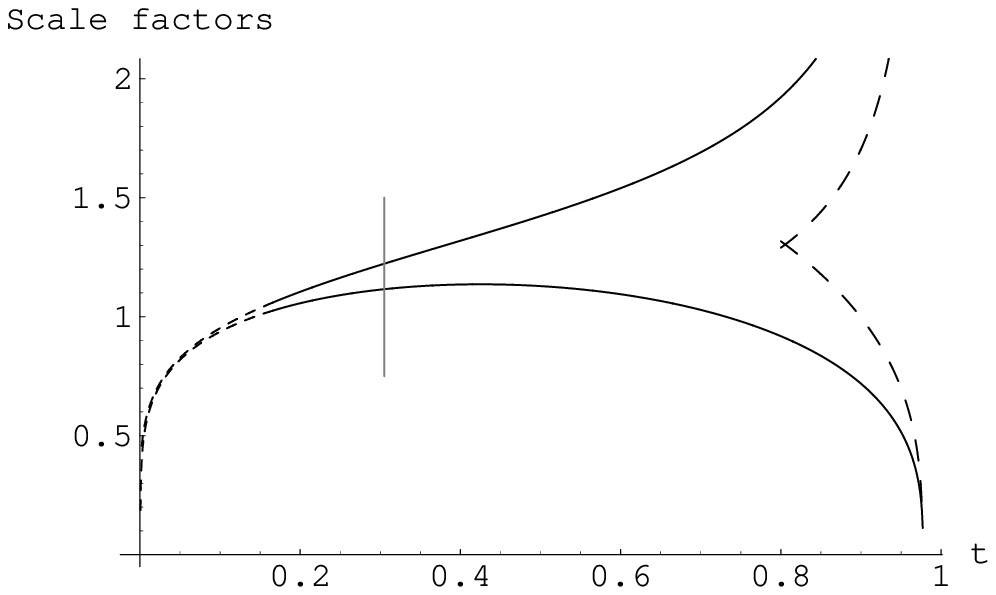}}
\end{figure}
\begin{figure}[t]
\caption{\label{fig:1d} Scale factors of the outer and inner
spaces in d.(210). The notation of the curves is the as in Fig. 2.} 
\centerline{\includegraphics[width=8cm]{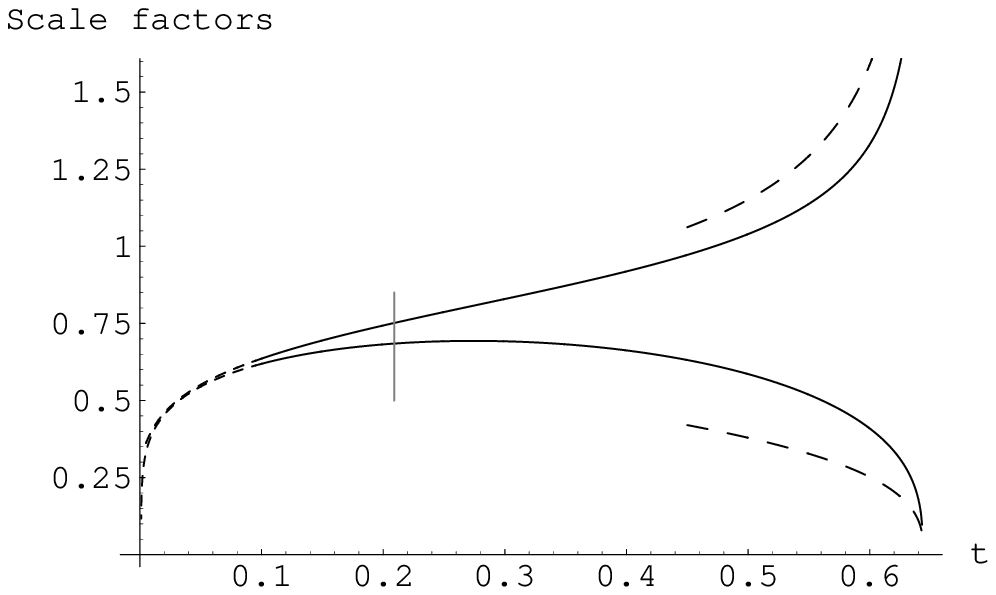}}
\end{figure} 
\begin{figure}[t]
\caption{\label{fig:1e} Scale factors of the outer and inner
spaces in e.(207). The notation of the curves is  as in Fig. 2.} 
\centerline{\includegraphics[width=8cm]{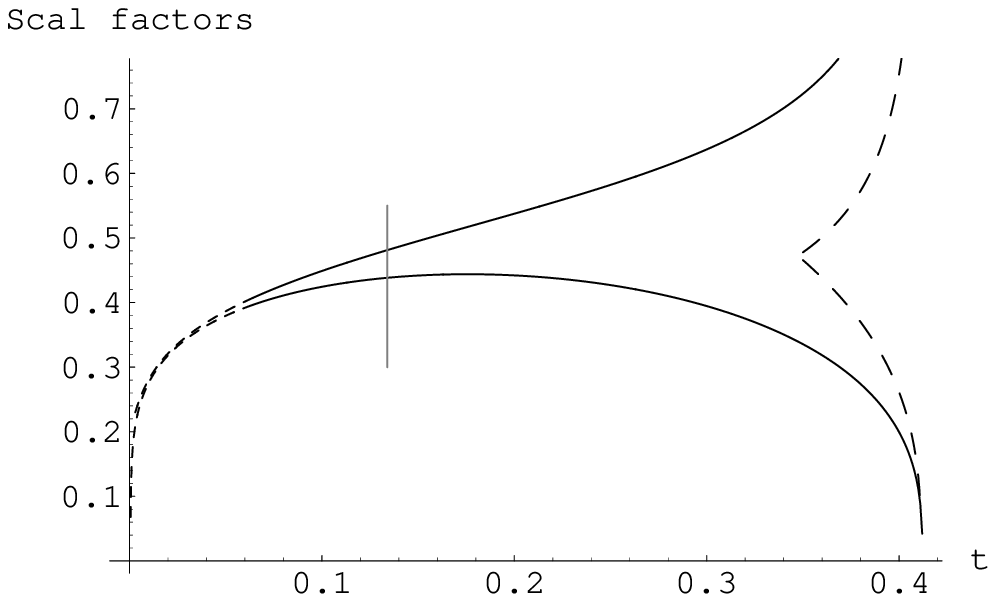}}
\end{figure}
\begin{figure}[t]
\caption{\label{fig:1f} Scale factors of the outer and inner
spaces in f.(215). The notation of the curves is  as in Fig. 2.} 
\centerline{\includegraphics[width=8cm]{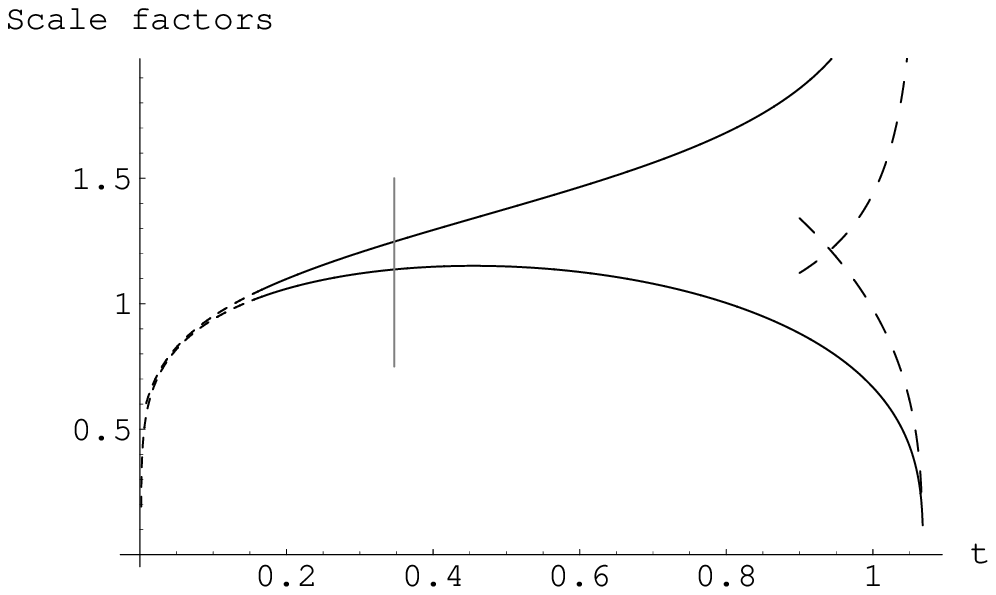}}
\end{figure}  
\begin{figure}[t]
\caption{\label{fig:2a} Factor $\Phi_0$ and temperature in a.(010). The
solid and dashed curves denote $\Phi_0$ and $T$, respectively.}
\centerline{\includegraphics[width=8cm]{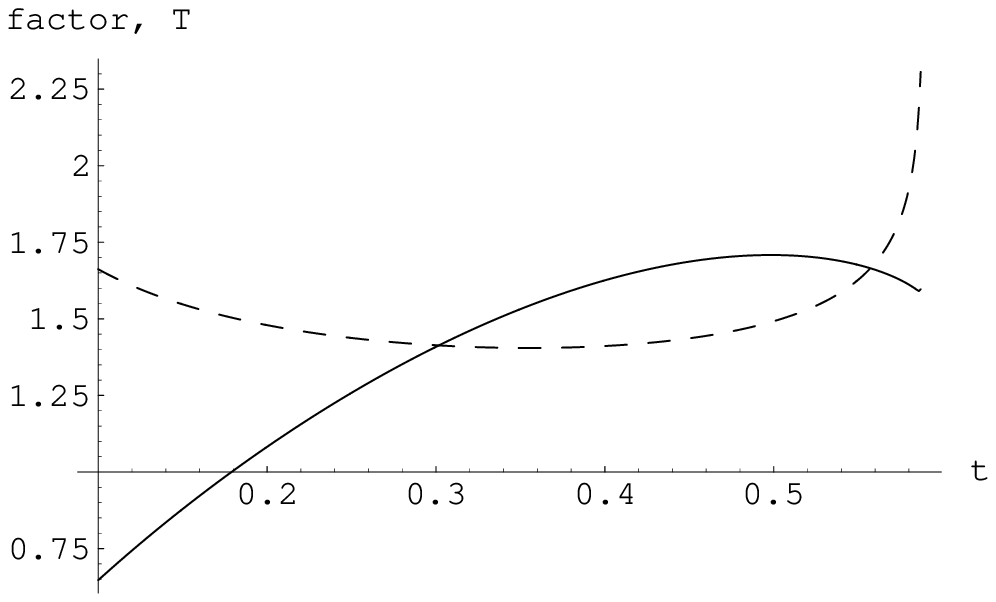}}
\end{figure}
\begin{figure}[t]
\caption{\label{fig:2b} Factor $\Phi_0$ and temperature in b.(007). The
notation of the curves is as in Fig. 8. }
\centerline{\includegraphics[width=8cm]{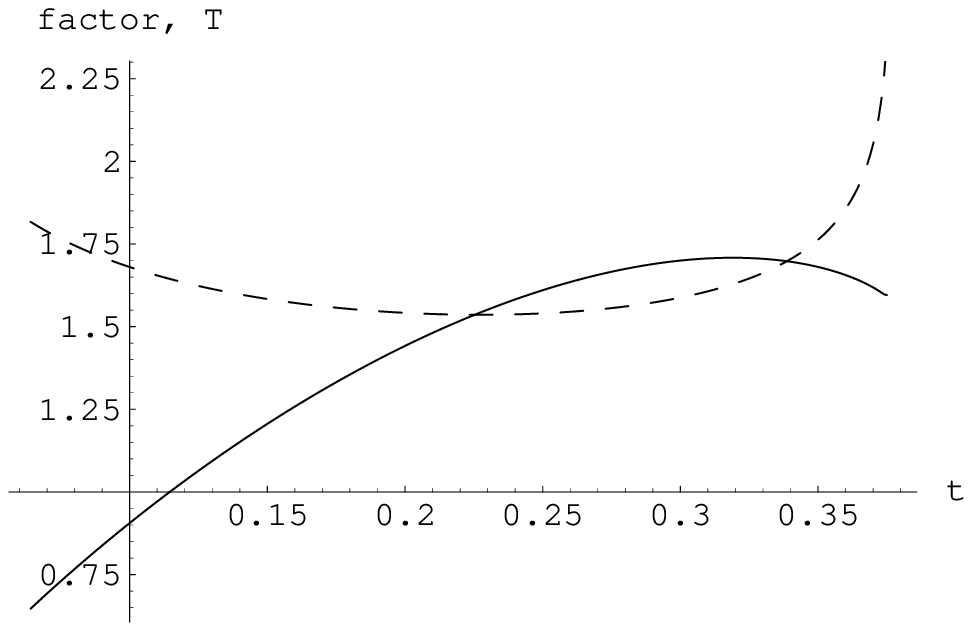}}
\end{figure}
\begin{figure}[t]
\caption{\label{fig:2c} Factor $\Phi_0$ and temperature in c.(015). The
notation of the curves is  as in Fig. 8. }
\centerline{\includegraphics[width=8cm]{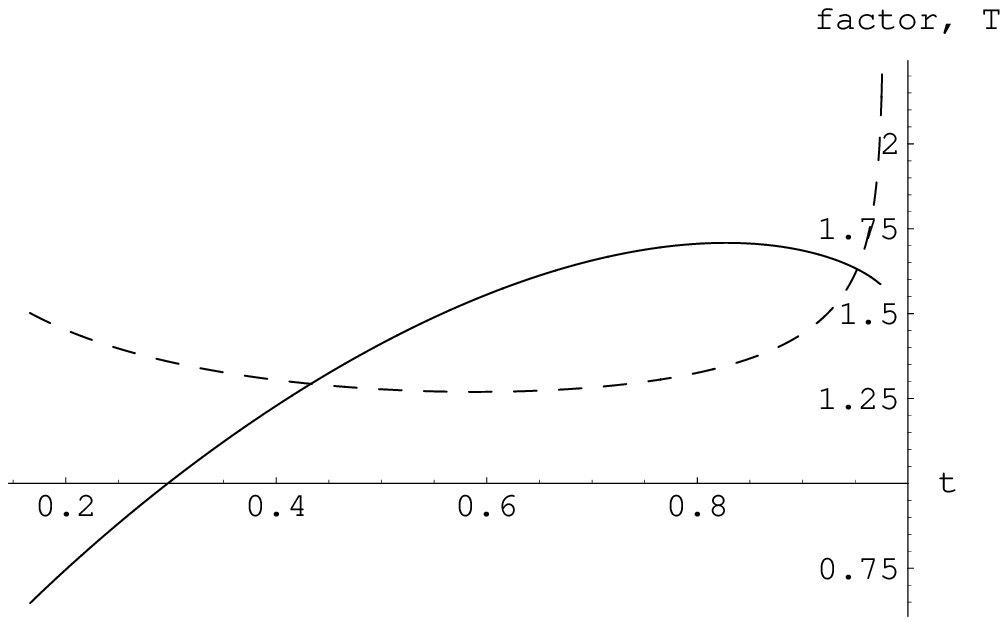}}
\end{figure}
\begin{figure}[t]
\caption{\label{fig:2d} Factor $\Phi_2$ and temperature in d.(210). The
solid and dashed curves denote $\Phi_2$ and $T$, respectively.}
\centerline{\includegraphics[width=8cm]{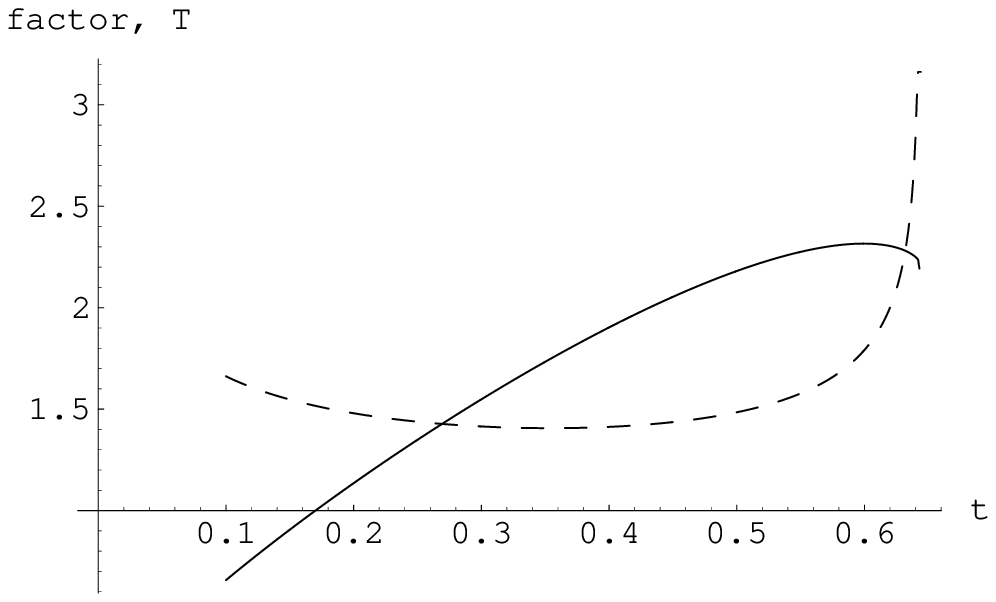}}
\end{figure}
\begin{figure}[t]
\caption{\label{fig:2e} Factor $\Phi_2$ and temperature in e.(207). The
notation of the curves is  as in Fig. 11.}
\centerline{\includegraphics[width=8cm]{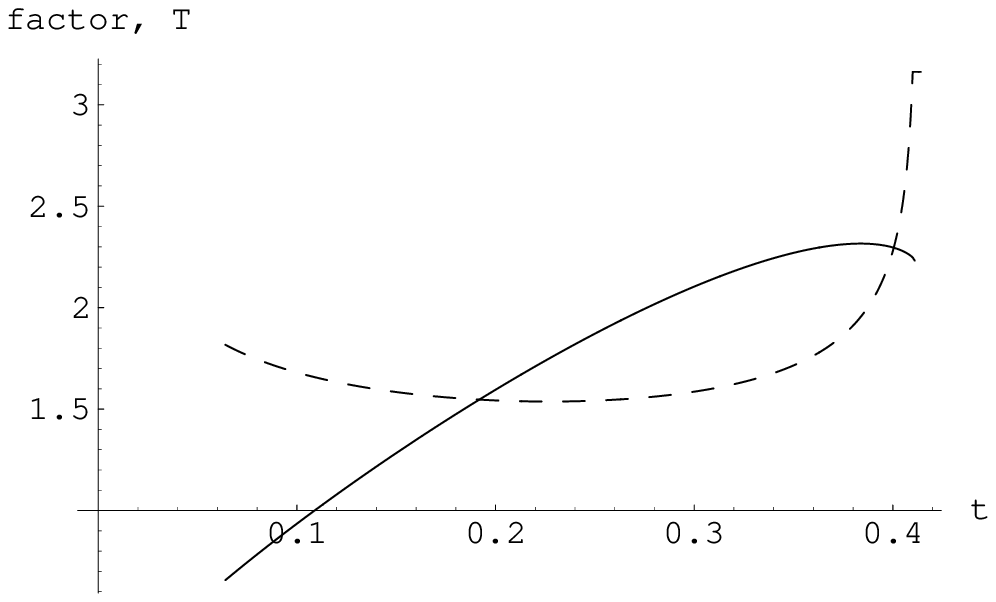}}
\end{figure}
\begin{figure}[t]
\caption{\label{fig:2f} Factor $\Phi_2$ and temperature in f.(215). The
notation of the curves is  as in Fig. 11.}
\centerline{\includegraphics[width=8cm]{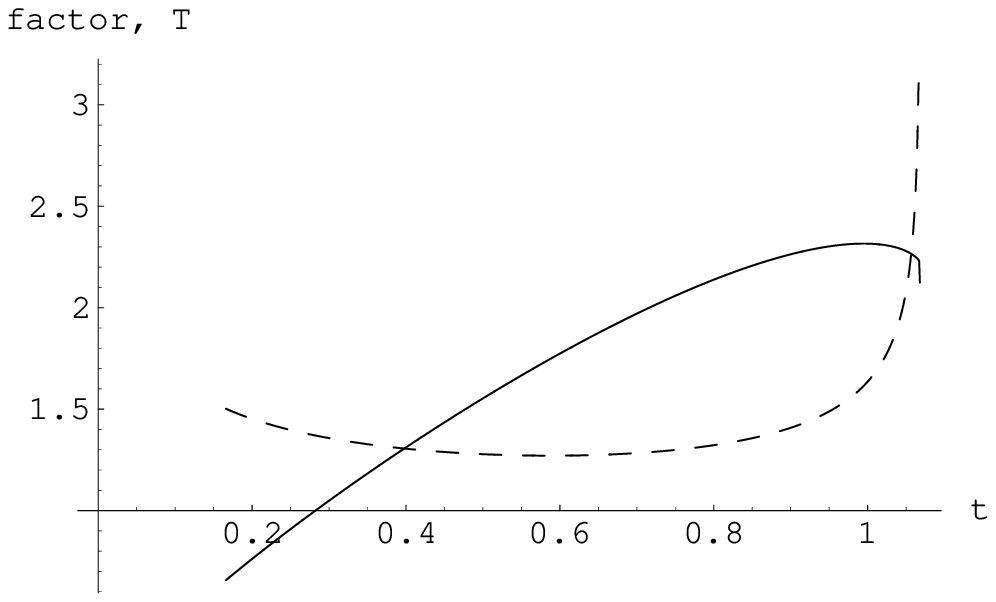}}
\end{figure}

\section{Physical states at the initial and decoupling epochs and the
primeval entropy} 
Let us use the formulation in Appendix B to
derive $\lambda \ [\equiv h(t_\dagger) T_\dagger]$ and the formulation
in Sect. 4 (B) to derive $S_0, \ h_M \ [\equiv \ h(t_M)]$ and $T_M \
[\equiv \ T(t_M)]$, where $t_M$ is the earliest epoch in the time
interval when each solution can be approximated by the corresponding
power solutions. For this purpose we first determine the epochs $t_A$
and $t_M$, and the factors $\Phi_0$ and $\Phi_2$ at epoch $t_M$, 
$(f/h)_M, h_i/h_M$ and $T_i/T_M$ using the
numerical results. Their result is shown in Tables 1 and 2 for $k_f = 0$ and
$-1$, respectively.

Next $\lambda, S_0, T_M$ and $h_M$ are obtained using the following
formulas (derived in Sect. 4) and the values of $\Phi_{0M}, \Phi_{2M}$ and 
$(f/h)_M$ (in Tables 1 and 2).  

\noindent Case of $\eta_0 = 0$
\begin{equation}
  \label{eq:bb1}
\begin{split}
\lambda &= 46.13 [\Phi_0 (t_M)]^{-0.1715} ({\cal N} a_n)^{-0.08576}, \\
S_0 &= \lambda^{7.5} {(f/h)_M}^3, \\
T_M/T_{pl} &= 10^{-4.89} \lambda (S_0)^{0.0556} , \\
h_M/f_{pl} &= 10^{14.7} {S_0}^{-0.167}.
\end{split}
\end{equation}
\noindent Case of $\eta_0 = 0.225$
\begin{equation}
  \label{eq:bb2}
\begin{split}
\lambda &= 1.164\times 10^4 [\Phi_2 (t_M)]^{-0.1307} ({\cal N} a_n)^{-0.06536}, \\
S_0 &= \lambda^{7.5} {(f/h)_M}^3, \\
T_M/T_{pl} &= 10^{-9.51} \lambda (S_0)^{0.108}, \\
h_M/f_{pl} &= 10^{16.4} {S_0}^{-0.186}.
\end{split}
\end{equation}
Moreover, $T_i$ and $h_i$ are derived from  $T_M$ and $h_M$ using
$(h_i/h_M)$ and $(T_i/T_M)$ (given in Tables 1 and 2).
In all cases a, ..., f, we show their values in Tables 3 and 4 for
$k_f = 0$ and $-1$, respectively, in the case of ${\cal
N}a_n = 1$. The values of $t_M$ were determined due to the comparison
between the numerical solutions ($f$ and $h$) and their corresponding power
 solutions. So, $h_M$
and $T_M$ depend on the epoch $t_M$ , but $h_i$ and $T_i$ do not depend on it.  

Here $f_i/h_i$ and $t_i/h_i$ are derived from the relations in
Sect. 5 as
\begin{equation}
  \label{eq:bb3}
\begin{split}
f_i/h_i &= (1 + f_{10} {t_i}^{8/5})/(1 + h_{10} {t_i}^{8/5}) \simeq
1+ 0.0327 (1 - 2k_f)/(1+\eta_0), \\
h_i/t_i &= h_0 {t_i}^{-4/5} (1+h_{10}{t_i}^{8/5}) \simeq
0.0251^{-1/2} = 6.325.  
\end{split}
\end{equation}

Now let us consider the total entropy $S_9$ within the volume 
$V (= f^3 h^6)$ at epoch $t_i$. From Eqs. (\ref{eq:a1}) and (\ref{eq:a4}), 
the total entropy $S$ is given by $S = \frac{5}{4} {\cal N} a_n V T^9$
 for $n = 6$, so that $S_9$ is defined by 
\begin{equation}
  \label{eq:bb4}
S_9 \equiv {T_i}^9 ({f_i}^3 {h_i}^6) = (T_i h_i)^9 (f_i/h_i)^3.
\end{equation}
The values in cases a, ..., f on the ordinary scale are shown in Table 5 for $k_f =
0$ and $-1$.  In the case of $k_f = 0$ the average values of
$S_9$ for $\eta_0 = 0$ and $0.225$ are 
$2.75\times 10^{90}$ and $1.29\times 10^{77}$, respectively,
and so $(S_9)_{\eta_0=0}/ (S_9)_{\eta_0=0.225} = 2.13\times
10^{13}$. In the case of $k_f = -1$ the average values of
$S_9$ for $\eta_0 = 0$ and $0.225$ are 
$2.87\times 10^{90}$ and $1.99\times 10^{77}$, respectively,
and so $(S_9)_{\eta_0=0}/ (S_9)_{\eta_0=0.225} = 1.45\times
10^{13}$.

As shown in Appendix C, $S_9$ and the $3$-dimensional entropy
$(S_3)_\dagger$ are closely related, and so these values of $S_9$ 
are the primeval total entropies necessary at the starting point
of multi-dimensional universes. In the non-viscous case ($\eta_0 =
0$), the entropy $S_9$ equal to $\sim 10^{90} [(S_3)_\dagger/
10^{88}]^2$ is needed, while in the viscous case of $\eta_0 = 0.225$,
 only the entropy $S_9$ equal to $\sim 10^{77} [(S_3)_\dagger/
10^{88}]^{0.71}$ is needed at the starting point, and so most entropy 
is produced by the dissipation.

We assumed ${\cal N} a_n = 1$ in the above calculations. This factor
${\cal N} a_n$ appears often in the calculations, as $h_i/f_{pl} 
\propto ({\cal N} a_n)^{0.1074}$ and $T_i/T_{pl} \propto  ({\cal N}
a_n)^{-0.1215}$ for $\eta_0 = 0$, and $h_i/f_{pl} 
\propto ({\cal N} a_n)^{0.09118}$ and $T_i/T_{pl} \propto  ({\cal N}
a_n)^{-0.1182}$ for $\eta_0 = 0.225$. But the product does not much
depend on ${\cal N} a_n$ as $T_i h_i \ \propto ({\cal N}
a_n)^{0.014}, \quad ({\cal N} a_n)^{0.027}$, respectively.   
 
\begin{table}
\caption{Model parameters in cases a, b, ..., f \ ($k_f = 0$). $\Phi_{0M} = 
\Phi_0(t_M)$ and $\Phi_{2M} = \Phi_2(t_M)$
\label{table1}}
\begin{tabular}{lcccccc}
model types
&$t_A$&$t_M$&$\Phi_{0M}, \ \Phi_{2M}$&$(f/h)_M$&$h_i/h_M$&$T_i/T_M$
\\ 
a. $(010)$\ \ & $0.589$ & $0.575$ & $1.63$ & $7.30$& $2.110$& $0.9092$\\
b. $(007)$\ \ & $0.376$ & $0.370$ & $1.62$ & $8.49$& $2.279$& $0.8855$\\ 
c. $(015)$\ \ & $0.977$ & $0.970$ & $1.58$ & $15.34$& $3.074$& $0.7996$\\
d. $(210)$\ \ & $0.644$ & $0.630$ & $2.30$ & $5.74$& $2.217$& $0.7500$\\ 
e. $(207)$\ \ & $0.412$ & $0.410$ & $2.25$ & $13.69$& $3.585$& $0.5746$\\ 
f. $(215)$\ \ & $1.068$ & $1.060$ & $2.26$ & $10.65$& $3.090$& $0.6222$\\ 
\end{tabular}
\end{table}
%
\begin{table}
\caption{Model parameters in cases a, b, ..., f \ ($k_f = -1$). $\Phi_{0M} = 
\Phi_0(t_M)$ and $\Phi_{2M} = \Phi_2(t_M)$
\label{table2}}
\begin{tabular}{lcccccc}
model types
&$t_A$&$t_M$&$\Phi_{0M}, \ \Phi_{2M}$&$(f/h)_M$&$h_i/h_M$&$T_i/T_M$
\\ 
a. $(010)$\ \ & $0.582$ & $0.570$ & $2.08$ & $10.05$& $2.321$& $0.9155$\\
b. $(007)$\ \ & $0.372$ & $0.366$ & $2.08$ & $11.14$& $2.445$& $0.8996$\\ 
c. $(015)$\ \ & $0.965$ & $0.958$ & $2.07$ & $19.62$& $3.251$& $0.8171$\\
d. $(210)$\ \ & $0.643$ & $0.630$ & $3.34$ & $7.92$& $2.318$& $0.7568$\\ 
e. $(207)$\ \ & $0.412$ & $0.410$ & $3.34$ & $19.41$& $3.893$& $0.5735$\\ 
f. $(215)$\ \ & $1.068$ & $1.060$ & $3.34$ & $14.56$& $3.299$& $0.6202$\\ 
\end{tabular}
\end{table}
\begin{table}
\caption{Model parameters in cases a, b, ..., f \ ($k_f = 0$).
\label{table3}}
\begin{tabular}{lcccccc}
$$ &$\lambda$&$S_0$&$10^3 \ T_M/T_{pl}$&$10^3 \ T_i/T_{pl}$&$h_M/f_{pl}
$&$h_i/f_{pl}$\\ 
a  & $42.42$ & $6.266\times 10^{14}$ & $3.437$ & $3.043$& $1.694\times 10^{12}$& $3.564\times
10^{12}$\\ 
b  & $42.47$ & $9.946\times 10^{14}$ & $3.532$ & $3.127$& $1.568\times 10^{12}$& $3.574\times
10^{12}$\\ 
c  & $42.65$ & $6.051\times 10^{15}$ & $3.924$ & $3.138$& $1.158\times 10^{12}$& $3.558\times
10^{12}$\\ 
d  & $1.044\times 10^4$ & $2.605\times 10^{32}$ & $9.998$ & $7.498$& $2.348\times 10^{10}$& $5.205\times 10^{10}$\\ 
e  & $1.047\times 10^4$ & $3.615\times 10^{33}$ & $13.27$ & $7.626$& $1.440\times 10^{10}$& $5.160\times 10^{10}$\\
f  & $1.046\times 10^4$ & $1.699\times 10^{33}$ & $12.26$ & $7.629$& $1.652\times 10^{10}$& $5.114\times 10^{10}$\\

\end{tabular}
\end{table}
\begin{table}
\caption{Model parameters in cases a, b, ..., f \ ($k_f = -1$).
\label{table4}}
\begin{tabular}{lcccccc}
$$ &$\lambda$&$S_0$&$10^3 \ T_M/T_{pl}$&$10^3 \ T_i/T_{pl}$&$h_M/f_{pl}$&$h_i/f_{pl}$
\\ 
a & $40.67$ & $1.191\times 10^{15}$ & $3.369$ & $3.084$& $1.522\times 10^{12}$& $3.533\times
10^{12}$\\ 
b & $40.68$ & $1.626\times 10^{15}$ & $3.429$ & $3.085$& $1.445\times 10^{12}$& $3.533\times
10^{12}$\\ 
c & $40.73$ & $8.968\times 10^{15}$ & $3.924$ & $3.085$& $1.086\times 10^{12}$& $3.533\times
10^{12}$\\ 
d & $0.9944\times 10^4$ & $4.771\times 10^{32}$ & $10.16$ & $7.693$& $2.098\times 10^{10}$& $4.864\times
10^{10}$\\ 
e & $0.9941\times 10^4$ & $6.996\times 10^{33}$ & $13.57$ & $7.782$& $1.273\times 10^{10}$& $4.966\times 10^{10}$\\
f & $0.9943\times 10^4$ & $2.958\times 10^{33}$ & $12.37$ & $7.672$& $1.474\times 10^{10}$& $4.863\times 10^{10}$\\

\end{tabular}
\end{table}
\begin{table}
\caption{The primeval total entropies $S_9$ in the cases of $k_f = \ 0, -1$. 
The values of $10^{-77} \ S_9$ are shown. 
\label{table5}}
\begin{tabular}{ccccccc}
model types &a&b&c&d&e&f\\
$10^{-77} S_9\ (k_f = \ 0)$\ & $2.29\times 10^{13}$& $3.00\times 10^{13}$ &
$2.97\times 10^{13}$ &\ $1.26$ &\ $1.36$&\ $1.26$\\
$10^{-77} S_9\ (k_f = -1)$\ & $2.87\times 10^{13}$& $2.88\times 10^{13}$ &
$2.88\times 10^{13}$ &\ $1.81$ &\ $2.38$&\ $1.77$\\  
\end{tabular}
\end{table}
%

\section{Dimensional symmetry-breaking and the primeval state of
multi-dimensional universes}
Kim et al.\citep{kim1,kim2} showed in a matrix model of super-string
theory that, due 
to the symmetry-breaking, the $(1+3+6)$-dimensional universe with
isotropic expansion changes to that with anisotropic expansion, in
which the $3$-dimensional space expands at the larger rate than the
$6$-dimensional space. In the present treatment due to classical
relativity, such a symmetry-breaking cannot be studied accurately.
From the viewpoint of energy balance, however, we may examine the behavior
of the symmetry breaking in simplified situations.

\medskip

\noindent 1. Case of $k_f = 0$
 
Let us assume that at epoch $t_{br}$ the symmetry-breaking occurred
from the isotropic state with scale factors $\bar{f} = \bar{h} \
(k_{\bar{f}} = k_{\bar{h}} = 1)$ to
the anisotropic state with $f \ne h \ (k_f = 0$ and $k_h = 1)$,
without change in $\dot{f}/f, \dot{h}/h$ and $\epsilon$. Then from 
Eq. (\ref{eq:a13}), we obtain the following
condition for consistency 
\begin{equation}
  \label{eq:cc1}
\frac{18}{{\bar{h}_{br}}^2} = \frac{15}{{h_{br}}^2} \quad {\rm or} \quad
h_{br}/\bar{h}_{br} = \sqrt{5/6}, 
\end{equation}
so that 
\begin{equation}
  \label{eq:cc2}
(\delta h/\bar{h})_{br} \equiv [(h-\bar{h})/\bar{h}]_{br} = 
 \sqrt{5/6} -1= -0.087.
\end{equation}
On the other hand, since we assume that $f$ does not change, i.e. $(\delta
f/\bar{f})_{br} = 0$, we have
\begin{equation}
  \label{eq:cc3}
(\delta h/\bar{h} - \delta f/\bar{f})_{br} = -0.087.
\end{equation}
This difference can be regarded as the difference of $f$ and $h$ from
the average at the instant of symmetry-breaking, which is given by
Eq.(\ref{eq:aa5a}) in Sect. 5:
\begin{equation}
  \label{eq:cc4}
(\delta h/\bar{h} - \delta f/\bar{f})_{br} = -\frac{125\times
3}{288}\frac{1}{1+\eta_0} {h_0}^{-2} (t_{br}/t_i)^{8/5} {t_i}^{8/5}. 
\end{equation}
From Eq.(\ref{eq:cc3}) and this relation we obtain
\begin{equation}
  \label{eq:cc5}
t_{br}/t_i = 1.84 (1+ \eta_0)^{5/8} = 1.84, \ \ 2.09
\end{equation}
for $\eta_0 = 0, \ \ 0.225$, respectively. These epochs ($t_{br}$) are
indicated in Figs. 2, ..., 7 by thin sold lines.

\medskip

\noindent 2. Case of $k_f = -1$

We assume also in this case that at epoch $t_{br}$ the 
symmetry-breaking occurred from the isotropic state with  
with scale factors $\bar{f} = \bar{h} \
(k_{\bar{f}} = k_{\bar{h}} = 1)$ to
the anisotropic state with $f \ne h \ (k_f = -1$ and $k_h = 1)$,
without change in $\dot{f}/f, \dot{h}/h, f$ and $\epsilon$. Then
similarly to the case of $k_f = 0$, we obtain the 
following condition for consistency from Eq. (\ref{eq:a13})
\begin{equation}
  \label{eq:cc6}
\frac{21}{{\bar{h}_{br}}^2} = \frac{15}{{h_{br}}^2} \quad {\rm or} \quad
h_{br}/\bar{h}_{br} = \sqrt{5/7},  
\end{equation}
so that 
\begin{equation}
  \label{eq:cc7}
(\delta h/\bar{h})_{br} \equiv [(h-\bar{h})/\bar{h}]_{br} = 
 \sqrt{5/7} -1= -0.155.
\end{equation}
On the other hand, since we assume that $f$ does not change, we have
\begin{equation}
  \label{eq:cc8}
(\delta h/\bar{h} - \delta f/\bar{f})_{br} = -0.155.
\end{equation}
This difference can be regarded as the difference of $f$ and $h$ from
the average at the instant of symmetry breaking, which is given by
Eq.(\ref{eq:aa5a}): 
\begin{equation}
  \label{eq:cc9}
(\delta h/\bar{h} - \delta f/\bar{f})_{br} = -\frac{125\times
9}{288}\frac{1}{1+\eta_0} {h_0}^{-2} (t_{br}/t_i)^{8/5} {t_i}^{8/5}. 
\end{equation}
From Eq.(\ref{eq:cc8}) and this relation we obtain
\begin{equation}
  \label{eq:cc10}
t_{br}/t_i = 1.57 (1+ \eta_0)^{5/8} = 1.57, \ \ 1.78
\end{equation}
for $\eta_0 = 0, 0.225$, respectively.

The primeval entropy $S_9$ within the closed $9$-dimensional space
before the symmetry-breaking is equal to $(S_9)_i \times
{(\bar{h}/h)_{br}}^6 = 1.73 (S_9)_i$, where $(S_9)_i$ is given by
Eq. (\ref{eq:bb4}).   

\section{Concluding remarks}
In this paper we first derived the $3$-dimensional entropy $S_3$ within 
the horizon, and compared it in the viscous and non-viscous cases. It
was found that the time evolutions of temperature $T$ and $S_3$ in the
viscous case are much larger than those in the non-viscous case. Such
a remarkable change in these quantities is caused by the change in
the energy density $\epsilon$ and the viscous dissipation. 
Next we derived the values of $f, h,$ and $T$ at the epoch ($t_\dagger$)
satisfying the condition that the 
$3$-dimensional entropy $S_3$ within the horizon should be $\sim
10^{88}$. Then, we examined the 
condition $(f/f_{pl})_\dagger \gg 1$, where $f_{pl}$ is the Planck
length in the outer space, and it was found that $\lambda (\equiv (h
T)_\dagger)$ must be $\gg 1.96$ and
$10^{-5.88}$ in the viscous and non-viscous cases,
respectively. It was found, moreover, that, at epoch $t_\dagger$
with $\lambda  \sim 1.06\times 10^4, 46.1$ (different from the
freeze-out epoch $t_*$), 
the condition $(h/f_{pl}) \sim 1$ is satisfied at the same time, so
that the outer space may be decoupled from the inner space. 

Moreover in Sect. 6 we considered the primeval entropy $S_9$,  
and found that, for an equal value of $S_3$,  $S_9$ in
viscous case can be much smaller than that in non-viscous case.
If $S_9$ in the universe at the primeval stage is comparable with the 
critical value  (such as in Table 5), the outer space does 
not need any additional $3$-dimensional inflation and dissipation after 
the decoupling epoch. If $S_9$ in the universe is much smaller than
the critical value, on the other hand, the outer space must create an additional
entropy by the 3-dimensional inflation and dissipation after the decoupling
 epoch, so that $S_3$ may reach $\sim 10^{88}$.  

Here we considered only the viscosity
due to the transport of multi-dimensional gravitational waves. If we
consider quantum particle creation\citep{maeda} and the other transport
processes, 
the viscous entropy production and $S_3$ may be much larger than those
in the present treatment.

Our results were derived using classical relativity and
thermodynamics. As the decoupling epoch may be in the world of quantum
super-string, the results may not hold in the original form, but
may give a qualitative trend of entropy production. 

\section*{Acknowledgments}
I like to express many thanks to Prof. H. Ishihara for helpful comments and
discussions. 
The numerical calculations in this work were carried out on SR16000 at
YITP in Kyoto University.

\appendix
\section{Imperfect fluid}
The energy-momentum tensor of an imperfect fluid is expressed as
\begin{equation}
  \label{eq:c1}
\begin{split}
T^{MN} &= \epsilon u^M u^N +\ p H^{MN} \  
-\ \eta H^{MK} H^{NL} [u_{K;L} +u_{L;K}
-\frac{2}{3+n}g_{KL}\Theta] \\
&- \ \zeta H^{MN} \Theta \ -\ \chi (H^{MK} u^N
+H^{NK} u^M) (T_{;K} +Tu_{K;L}u^L),  
\end{split}
\end{equation}
where the $(4+n)$-dimensional velocity $u^M$ satisfies $g_{MN} u^M u^N =
-1$, and $H^{MN}$ is the projection tensor defined by
\begin{equation}
  \label{eq:c2}
H^{MN} = g^{MN} + u^M u^N,
\end{equation}
and $\Theta$ denotes ${u^L}_{;L}$. From the equation of relativistic
radiative transport, we obtain (see Appendix B of Ref.\citep{TI}
for details) :
\begin{equation}
  \label{eq:c3}
\begin{split}
\eta &= \frac{4+n}{(3+n)(5+n)} \epsilon_r \tau, \\
\zeta &= (4+n) \epsilon_r \tau [1/(3+n) -(\partial
p/\partial\epsilon)_V ]^2, \\
\chi &= \frac{(4+n)}{(3+n)} \epsilon_r \tau, \quad  \epsilon_r = {\cal
N}_r a_n T^{4+n},
\end{split}
\end{equation}
where $\eta, \zeta,$ and $\chi$ are the shear viscosity, the bulk
viscosity, and the heat conductivity, respectively, $\tau$ is the mean
free time and ${\cal N}_r$ is the number of 
radiative particle species.  For the radiative matter,
we have $\zeta = 0$ and $\chi$ does not contribute to our present treatment.
A misprint was found in the expression for $\eta$ in the previous 
paper\citep{TI}.

The gravitational waves are absorbed in imperfect fluids. Their mean
free time $\tau$ is expressed as 
\begin{equation}
  \label{eq:c4}
\tau = (2\kappa\eta)^{-1},
\end{equation}
as shown in Appendix A of Ref.\citep{TI}.

\section{The Planck length $f_{pl}$ in the outer space}
Let us consider the relation of $f$ and $h$ with the Planck
length $f_{pl}$ given by Eq.(\ref{eq:d1}).

\noindent (a) the viscous case

From Eq.(\ref{eq:d1}) we get 
\begin{equation}
  \label{eq:e1}
{\Bigl(\frac{f}{f_{pl}}\Bigr)_\dagger}^2 = \Bigl[\frac{8\pi}{
\kappa \epsilon} (f^2/h^4) \Bigr]_M {\lambda}^{4+n} 
 \Bigl[\frac{f_\dagger {h_\dagger}^{-2} (t_A-t_\dagger)}{f_M {h_M}^{-2}
(t_A-t_M)} \Bigr]^2 {\cal N} a_n, 
\end{equation}
where $\kappa \epsilon$ is given by  Eqs.(\ref{eq:a13}) and
(\ref{eq:a22}). Here we use Eqs.(\ref{eq:b17}) and (\ref{eq:b19}) and
the $t$-dependence of $f$ and $h$, such as $fh^{-2} (t_A - t) \propto
h^{-2 + (\nu +1)/\mu}$.  Then  we
obtain for $n = 6$
\begin{equation}
  \label{eq:e2}
\Bigl(\frac{f}{f_{pl}}\Bigr)_\dagger = 10^{-1.804}
\Bigl[\sqrt{\frac{8\pi}{\kappa \epsilon}} (f/h^2) \Bigr]_M
\lambda^{5.15} {(f/h)_M}^{0.06} ({\cal N} a_n)^{1/2}
\end{equation}
and 
\begin{equation}
  \label{eq:e3}
\Bigl(\frac{h}{f_{pl}} \Bigr)_\dagger = \Bigl(\frac{h}{f} \Bigr)_\dagger
\Bigl(\frac{f}{f_{pl}} \Bigr)_\dagger = 10^{-29.3} \lambda^{5/2}
\Bigl(\frac{f}{f_{pl}} \Bigr)_\dagger. 
\end{equation}
Here we estimate $\Bigl[\sqrt{\frac{8\pi}{\kappa \epsilon}} (f/h^2)
\Bigr]_M {(f/h)_M}^{0.06}$ \ ($\equiv \Phi_2 (t_M)$) to be $2$, since $t_M$ is
 near the epoch of maximum expansion of the inner space. Its 
 model-dependent value ($\Phi_2$) is shown in Sect. 6.
Then if we assume $(f/f_{pl})_\dagger \gg 1$, we
get the condition $ \lambda \gg 1.96$. But if $\lambda < 1.06\times 10^4$,
$h/f_{pl}$ is smaller than $1$.  Here the factor ${\cal N} a_n$ was
neglected, because its contribution is small. 
 
\noindent (b) the non-viscous case

From Eq.(\ref{eq:d1}) we get, similarly,
\begin{equation}
  \label{eq:e4}
{\Bigl(\frac{f}{f_{pl}}\Bigr)_\dagger}^2 = \Bigl[\frac{8\pi}{
\kappa \epsilon} (f^2/h^4) \Bigr]_M {\lambda}^{4+n} \Bigl[\frac{f_\dagger
{h_\dagger}^{-2} (t_A-t_\dagger)^{\alpha/2}}{f_M {h_M}^{-2}
(t_A-t_M)^{\alpha/2}} \Bigr]^2 {\cal N} a_n, 
\end{equation}
where $\kappa \epsilon$ is given by  Eqs.(\ref{eq:a13}) and
(\ref{eq:a22a}) and $\alpha = 10/9$ for $n=6$. 
Here we use Eqs.(\ref{eq:b17}) and (\ref{eq:b19b}) and
the $t$-dependence of $f$ and $h$, such as $fh^{-2} (t_A - t)^{\alpha/2} 
\propto h^{-2 + (\nu +\alpha/2)/\mu}$. Then we
obtain 
\begin{equation}
  \label{eq:e5}
\Bigl(\frac{f}{f_{pl}}\Bigr)_\dagger = 10^{19.6}
\Bigl[\sqrt{\frac{8\pi}{\kappa \epsilon}} (f/h^2)\Bigr]_M
\lambda^{3.33} {(h/f)_M}^{0.667} 
({\cal N} a_n)^{1/2} 
\end{equation}
and 
\begin{equation}
  \label{eq:e6}
\Bigl(\frac{h}{f_{pl}} \Bigr)_\dagger = \Bigl(\frac{h}{f} \Bigr)_\dagger
\Bigl(\frac{f}{f_{pl}} \Bigr)_\dagger = 10^{-29.3} \lambda^{5/2}
\Bigl(\frac{f}{f_{pl}} \Bigr)_\dagger. 
\end{equation}
Here we estimate $\Bigl[\sqrt{\frac{8\pi}{\kappa
\epsilon}} (f/h^2)\Bigr]_M 
{(h/f)_M}^{0.667}$ \ ($\equiv \Phi_0 (t_M)$) to be $1$, since $t_M$ is near the
epoch of maximum expansion of the inner space. Its model-dependent
 value ($\Phi_0$) 
is shown in Sect. 6. Then if we assume $(f/f_{pl})_\dagger \gg 1$, we
get the condition $ \lambda \gg 10^{-5.88}$. But if $\lambda < 46.1$,
$h/f_{pl}$ is smaller than $1$. On the other hand, $h/f_{pl} =
10^{-9.7}$ for $\lambda = 1$ (or at epoch $t_*$), which is consistent
with (5.7) of Kolb et al.\citep{kolb}.   
 
Thus, if  $\lambda (=(h T)_\dagger)$ is comparable with $1$,
$h/f_{pl}$ is much smaller than $1$ in both cases, and for $\eta_0 = 
(0.225, 0)$, we obtain  $(h/ f_{pl}) \sim 1$ if $\lambda \sim 
(1.06\times 10^4, 46.1)$.

\section{Dependence of $(S_9)_i$ on the $3-$dimensional entropy
$(S_3)_\dagger$ }
In the text we treated only the case when $(S_3)_\dagger =
10^{88}$. Here we consider the dependence of $(S_9)_i$ on
$(S_3)_\dagger/10^{88} \ \ (\equiv R)$, under the condition that
$h(t_\dagger)$ is equal to the Planck length $f_{pl}$.

\noindent (a) the viscous case

From Eqs.(\ref{eq:b19}), we obtain 
\begin{equation}
  \label{eq:f1}
\lambda/(hT)_M \propto [(S_3)/S_0]^{1/\delta},
\end{equation}
where $\delta = -12.77$  and $S_0 \equiv \lambda^{15/2} {(f/h)_M}^3$.
Here $\lambda$ is determined by the above condition (that
$h(t_\dagger)$ is equal to the Planck length) which is given in
Appendix B, but it is independent of
$S_3$. Therefore we have 
\begin{equation}
  \label{eq:f2}
(h T)_M \propto R^{0.0783},
\end{equation}
and using the ratios $h_i/h_M$ and $T_i/T_M$ in Table 1 and Table 2 and the
values of $\mu$ and $\nu$ (which are independent of $S_3$), we obtain
\begin{equation}
  \label{eq:f3}
T_i h_i \propto R^{0.0783},
\end{equation}
so that
\begin{equation}
  \label{eq:f4}
(S_9)_i = (T_i h_i)^9 \propto R^{0.7047}.
\end{equation}

\noindent (b) the non-viscous case

From Eqs.(\ref{eq:b19b}), similarly, we obtain 
\begin{equation}
  \label{eq:f5}
\lambda/(hT)_M \propto [(S_3)/S_0]^{-1/9},
\end{equation}
\begin{equation}
  \label{eq:f6}
T_i h_i \propto R^{1/9},
\end{equation}
so that
\begin{equation}
  \label{eq:f7}
(S_9)_i = (T_i h_i)^9 \propto R.
\end{equation}
It is found, therefore, that $(S_9)_i$ is proportional to $(S_3)_\dagger/10^{88}
 \ \ (\equiv R)$ for $\eta_0 = 0$ and the dependence of $(S_9)_i$ on
$R$ is smaller for $\eta_0 = 0.225$ than for $\eta_0 = 0$.


\end{document}